\documentstyle[multicol,aps,prl,epsfig]{revtex}
\begin{document}
\draft
\title{
CO oxidation on a single Pd atom supported on magnesia} 
\author{  
S. Abbet,$^1$ U. Heiz,$^2$
H. H\"akkinen$^3$  and U. Landman$^3$\\
$^1$Universit\'e de Lausanne, Institut de Physique de la Mati\`ere
Condens\`ee, CH-1015 Lausanne, Switzerland  \\
$^2$University of Ulm, 
Institute of Surface Chemistry and Catalysis, 89069 Ulm, Germany\\
$^3$School of Physics, Georgia Institute of Technology,
Atlanta, GA 30332-0430}
\date{to be published in PRL 18 June 2001} 
\maketitle
%
%
\begin{abstract}

The oxidation of CO on single Pd atoms anchored to
MgO(100) surface oxygen vacancies is studied with
temperature-programmed-reaction mass-spectrometry and
infrared spectroscopy. In one-heating-cycle experiments
CO$_2$, formed from O$_2$ and CO preadsorbed at 90 K,
is detected at 260 K and 500 K. Ab-initio 
simulations suggest two reaction routes, with 
Pd(CO)$_2$O$_2$ and Pd(CO$_3$)CO found as
precursors for the low and
high temperature channels, respectively. Both
reactions result in annealing of the vacancy and induce
migration and coalescence of the remaining Pd-CO
to form larger clusters. 

\end{abstract} 
\begin{multicols}{2}
\narrowtext
%
%
%
%
%

Molecular-scale understanding of the energetics and
mechanisms of catalytic reactions
could open new avenues to 
the design of catalysts with specific functions \cite{Bes98,Ertl99}.
To this aim model 
catalysts are used \cite{Ertl99} to extract information on reactivities
at conditions relevant to catalysis \cite{Somorjai,Henry,Freund}.
 A most fruitful approach to gain insights into the 
reaction mechanisms of catalytic processes
 is the combination of experiments and high-level atomic scale 
simulations.
Indeed, such recent joint  studies 
revealed the reaction mechanism of the oxidation of CO on Au$_8$ clusters
\cite{hak99}  and the 
polymerization of acetylene on Pd$_n$ ($n\leq 30$) clusters
deposited on MgO(100) films
\cite{jacs00}.

Here we report on studies of the oxidation of CO by a
model catalyst consisting of single Pd atoms anchored on oxygen 
surface vacancies (F-centers, FCs)
of a MgO(100) film. 
After low-temperature  (90 K)
coadsorption of O$_2$ and CO the formation of CO$_2$ was detected
by temperature-programmed-reaction (TPR) mass-spectrometry 
in  single-heating-cycle experiments. 
Ab-initio 
density-functional simulations were performed 
to identify relevant molecular 
precursors 
as well as to 
study the CO$_2$ formation mechanisms. 
The determination of two 
initial molecular complexes adsorbed on
MgO(100), Pd(CO)$_2$O$_2$ and 
Pd(CO$_3$)CO, is
supported by  good agreement between the measured and 
calculated CO vibrational frequencies. 
The former complex is involved in the formation 
of CO$_2$ at 260 K, and the decomposition 
of the carbonate complex leads to  CO$_2$ desorption
 at 500  
K. Both reaction routes induce annealing of the
surface FC and  migration of the remaining 
Pd-CO unit to form larger Pd clusters.       
 
{\it Model Catalyst Preparation and Experimental Results.}
Size-selected atomic 
Pd cations were deposited on an in-situ prepared MgO(100) thin 
film \cite{Goodman}. The low-kinetic-energy (0.2 eV)
deposition of only a 0.45 x $10^{-2}$ of a monolayer (ML) of Pd
 at a substrate temperature of 90  K reduces greatly the migration of the 
Pd atoms bound to the FCs (whose concentration is 10$^{-2}$ ML, i.e.,
roughly twice that of 
the adsorbed Pd atoms).
 Indeed,  a recent comparison of ab-initio calculations and 
FTIR studies of CO adsorbed on the supported Pd atoms 
provided clear evidence for   single Pd atoms 
bound to the FCs of the MgO(100) support.
The observation of two different 
vibrational bands at 90 K suggests the presence of at least two different CO 
molecules adsorbed on monodispersed Pd atoms (see ref. \onlinecite{defects} and
below).   A recent study 
on the cyclotrimerization of acetylene also 
revealed that the FC-trapped Pd atoms are stable up to 300 K \cite{jacs00}.    

First, we verified that the clean 
MgO(100) thin films are inert for the oxidation reaction; 
i.e., no CO$_2$ was formed in a 
one-heating-cycle experiment after adsorbing O$_2$ and CO or vice versa
\cite{heiz99}. When Pd
atoms are trapped on the FCs,
preadsorption of oxygen and subsequent saturation by 
CO leads to the formation of carbon dioxide, with desorption peaks
 at 260 K and at around 500 K (Fig. 1). 
The existence 
of two desorption peaks suggests the presence of two different reaction 
mechanisms. Note that
when  CO is preadsorbed prior to O$_2$ 
the oxidation reaction is suppressed, 
 indicating CO poisoning.

Information pertaining to  the 
mechanism of the CO oxidation on FC-trapped Pd atoms was  
obtained by measuring the CO 
vibrational bands during reaction (See insets to Fig. 1). At 
95 K three features in the FTIR 
spectrum of $^{13}$CO are observed. The infrared absorption 
at 2125 cm$^{-1}$ originates 
from CO adsorbed on extended defect sites on the MgO(100) thin 
films \cite{pacchioni}.
 The broad band with a peak 
at around 2045 cm$^{-1}$ and a shoulder at 2005 cm$^{-1}$ 
indicates adsorption of at least 
two CO molecules. Heating to 165 K results in a band 
at 2005 cm$^{-1}$ with a shoulder at 2035 cm$^{-1}$. 
The decreased intensity of the high-frequency 
band suggests partial desorption of the 
high frequency CO molecule prior to the oxidation 
reaction. At 250 K, close to the temperature 
of maximum CO$_2$ desorption, the shoulder at 
2035 cm$^{-1}$ almost disappears,  
indicating oxidation of the 
 $^{13}$CO molecule
with a vibrational frequency of 2035 cm$^{-1}$. 
 Further heating  to 410 K results in 
disappearance of the band at 2005 cm$^{-1}$. 
This  correlates with the formation of CO$_2$ and 
the observed desorption of molecular CO (not shown). 
In addition, 
 a new vibrational band with a 
frequency of 1830 cm$^{-1}$ appears
between 250 and 300 K. The   disappearance of this band
above 600 K correlates with  complete CO oxidation and molecular CO 
desorption.

{\it Theoretical Methodology and Results.}     
The calculations were 
performed using the Born-Oppenheimer (BO) local-spin-density 
(LSD) molecular dynamics method (BO-LSD-MD) \cite{BL} 
with the generalized gradient 
approximation (GGA) \cite{GGA} and employing 
norm-conserving non-local scalar-relativistic\cite{rel} (for the Pd atom) 
pseudopotentials \cite{TM}. 
Such calculations
 yield accurate results pertaining to
geometries, electronic structure, and charging effects
of various 
neutral and charged  
coinage metal clusters \cite{hak00} and
  nanostructures \cite{wire}.
The magnesia  surface was modeled by  a finite region ("cluster")
of atoms, whose valence electrons are treated fully quantum-mechanically  
(using the BO-LSD-MD), embedded  in a large (2000 charges) 
point-charge lattice, as 
described   
in  our study of the  Au$_8$/MgO model catalyst \cite{hak99}.

A single Pd atom
binds  strongly to the oxygen vacancy   
(binding energy of 3.31 eV),
with a slight amount of charge (0.15 e)
transferred to the adsorbed atom.
 In comparison, the binding energy of Pd 
atoms to terrace oxygen sites is only 1.16 eV. 
The enhanced binding to the FCs
is also reflected in the corresponding 
bonding lengths of 1.65 \AA\  and 2.17 \AA\ for MgO(FC)-Pd
and MgO-Pd, respectively \cite{Note2}. 

Binding of two CO molecules saturates   
the MgO(FC)-Pd system; 
occupying the MgO(FC)-Pd system with three CO molecules
leads to spontaneous (barrierless) desorption of
one of the molecules.
  In the most 
stable configuration 
the two CO molecules are inequivalent;
one CO binds 
 on-top 
 and the second  adsorbs on the side of the Pd-atom 
[9b] (this top-side
geometry is similar to that shown in
Fig. 2a but without the O$_2$), and 
 the total binding energy of the two CO molecules is 1.62 eV. An  alternative 
symmetric adsorption configuration,
 with the two CO molecules adsorbing on opposing 
sides of the Pd atom,  is  less stable by 0.61 eV than 
the top-side one [9b]. 
 
To study the oxidation mechanisms of CO on  MgO(FC)-Pd 
the system was optimized first 
with coadsorbed O$_2$ and two CO molecules.  Two stable geometric 
arrangements were found, with the most stable one
shown  in Fig. 2a where  the CO molecules bind  in a top-side 
configuration and the O$_2$ is adsorbed
 parallel to the surface on the other side of the Pd atom. 
This configuration (with spin $S=0$) is 0.90 eV 
more stable than an alternative one ($S=1$) where the O$_2$ is 
bound on-top of the Pd atom and the two
CO molecules occupy the side positions (not shown). The 
preadsorbed O$_2$ molecule enhances slightly the 
adsorption energy of the two CO molecules (1.78 
eV) compared to the case without preadsorbed oxygen  (1.62 eV). 
The adsorbed  O$_2$ molecular bond is stretched and activated
(1.46 \AA\ compared to the calculated gas-phase value of 1.25 \AA).
In addition,  we found a  stable
carbonate complex Pd(CO$_3$)CO (Fig. 3a), whose binding energy is
4.08 eV larger  than the aforementioned Pd(CO)$_2$O$_2$ complex.

{\it Reaction mechanisms.} 
Two  reaction mechanisms are proposed corresponding to  
the two CO$_2$  peaks observed 
experimentally (Fig. 1). At low temperatures the
two relevant 
precursors are shown in Fig. 2a and Fig. 3a.
The existence of the 
Pd(CO)$_2$O$_2$ complex (Fig. 2a) is supported by the agreement between the 
calculated  and
measured   vibrational 
frequencies of the two inequivalently  
adsorbed CO molecules (see Table I).
The absence of a clear FTIR signal corresponding to
the carbonate species can originate from the
adsorption geometry (see Fig. 3a)
where the carbonate is bound at the side of the Pd atom
and the
CO$_3$ plane is only slightly tilted away from the surface,
resulting in vanishingly small normal dynamic dipole
components; however, the
 frequency of the side-bonded CO of this
complex (2020 cm$^{-1}$) lies in the 
experimentally observed vibrational band. 

Corresponding to
 the 260 K CO$_2$ desorption peak we propose the following reaction 
mechanism. First, in a competitive process,
 CO desorbs or is oxidized upon heating. The 
theoretically estimated activation energies of the 
two processes are   0.89 eV for desorption (Table I)
and 0.84 for oxidation, obtained from a series of
constrained energy-minimizations, where the top-CO molecule
approaches the closest O atom of the O$_2$ molecule.
The transition  state leading to
 CO$_2$ 
formation  is shown in Fig. 2b. 
After finding the transition state we
performed a microcanonical MD simulation
to study the reaction dynamics.
 The desorbing CO$_2$ molecule  (Fig. 2c)
carries away the major part ($\sim$2 eV) 
of the  reaction heat of about 2.2 eV, partitioned as
0.1 eV, 0.1 eV, and 1.8 eV into the 
translational, rotational and vibrational degrees
of freedom, respectively, i.e., the desorbing molecule
is vibrationally "hot". 
The remaining O atom 
of the complex  fills the O vacancy under the adsorbed
 Pd atom (Fig. 2c), releasing 2.8 eV (not shown in Fig. 2d). 
Concomitantly with the annealing of the O vacancy
 the binding energy of the Pd atom is 
largely reduced 
from 3.31 to 1.16 eV, as discussed above.
 Note also that the binding energy of 
the remaining CO molecule to the Pd atom 
increases (the calculated binding energy of a CO
molecule to a Pd atom adsorbed on the  terrace  of MgO(100)
is 2.29 eV). Consequently, migration of the 
 Pd-CO unit, leading to formation of larger clusters,
 becomes energetically feasible 
(the calculated 
diffusion barrier for a Pd atom on the MgO(100)
terrace is 0.43 eV). Indeed, the observed vibrational 
band with a frequency of 1870 cm$^{-1}$ appearing 
between 250 K and 300 K (Fig. 1) is in close 
agreement with the calculated frequency of a CO molecule bridge-bonded
to an adsorbed Pd$_2$ dimer; for a Pd$_2$ adsorbed on a MgO(100)
surface the calculated frequency is 1836 cm$^{-1}$, and for a
Pd$_2$ adsorbed at an FC it is 1877 cm$^{-1}$ (Table I).
Desorption of
the adsorbed CO molecule occurs at temperatures below 600 
K. The calculated 
binding energy of the bridge-bonded CO to the dimer is 1.95 eV when 
it is adsorbed at an  FC and 2.87 eV 
when adsorbed on a terrace,  indicating that the 
Pd$_2$ dimer (with a bridge-bonded CO) is 
most likely to be located at an FC, 
 as otherwise an 
even higher CO desorption temperature 
is expected. Such FCs are available since the density of 
the oxygen vacancies  in our experiment is 
roughly twice that of the Pd atoms. 
   
Formation of CO$_2$ at higher 
temperatures (corresponding to desorption 
around 500K, Fig. 1) 
involves decomposition of
the Pd(CO$_3$)CO carbonate complex (Fig. 3a). This mechanism is 
 observed in molecular dynamics simulations where the 
temperature is controlled  
to 500 K by Langevin dynamics.  
 After the transition state (see Fig. 3b),
overcoming an energy barrier of about 1 eV (see Fig. 3d),
CO$_2$ 
leaves the complex parallel to the MgO surface with a
total kinetic energy of about 0.25 eV, distributed
approximately as 0.1 eV and
0.15 eV   between the translational and vibrational
modes, respectively, and with a vanishing rotational component.
The remaining O atom fills 
the O vacancy 
(Fig. 3c) as found also for the lower-temperature CO$_2$ formation mechanism.
 The total exothermicity of this process is 1.8 eV (see
the sharp drop in Fig. 3d for $t\geq 210$ fs). As in the 
low-temperature mechanism, the remaining
 Pd-CO  can migrate and coalesce to larger 
clusters. 

Finally, we remark that the relative abundance of the two
surface precursors, Pd(CO)$_2$O$_2$ and Pd(CO$_3$)CO, 
underlying the above reaction mechanisms
may be dominated by kinetic factors; e.g., 
formation of the carbonate complex,
although energetically favorable, could be
 hindered by the requirement that the CO molecule
will approach the   
preadsorbed side-O$_2$ 
rather than the Pd atom, 
as well as by a significant reaction barrier.
Further  studies of the low temperature formation of the carbonate 
complex, as well as  efforts to
identify such species (e.g. via HREELS),  are  warranted.

This research is supported by the 
Swiss National Science Foundation (U.H. and S.A.), by the US 
AFOSR (U.L. and H.H.), and by the Academy of Finland (H.H.). We thank W.-D. 
Schneider for his support and 
M. Moseler for fruitful discussions. The computations were 
performed on an IBM SP 
and a Cray T3E at the Center for Scientific Computing in 
Espoo, Finland.

%
%

\newpage

\begin{table}
\caption{Calculated CO binding energies ($E_B$)
 and $^{12}$CO vibrational
frequencies ($\omega^{th}$), compared to the experimental frequencies
($\omega^{exp}$,
scaled for $^{12}$CO
from the values shown in Fig. 1 for  $^{13}$CO).
 The calculated dissociation energy,
equilibrium bond length,
and harmonic frequency of the 
gas-phase $^{12}$CO molecule  
are 11.06 eV, 1.141 \AA, and 2140 cm$^{-1}$, respectively,
compared to the experimental values of 11.09 eV, 1.128 \AA, and
2170 cm$^{-1}$. The CO vibrational frequencies
are determined through molecular dynamics simulations
stretching the equilibrium CO bond by 1\% and
 observing the dynamics over a few  harmonic vibration periods.
 All the calculated $\omega$ values 
include a correction factor of 2170/2140. The error estimate for the
calculated frequencies is $\pm 10$ cm$^{-1}$. 
}

\label{tab1}
\begin{tabular}{llll}   
Complex  & $E_B$ (eV) 
& $\omega^{th}$ (cm$^{-1}$)  &  $\omega^{exp}$ (cm$^{-1}$)
\\
\tableline
MgO(FC)-Pd-(CO)$_2$-O$_2$  & 1.78\tablenote{per 2 CO molecules} & 
2019 / 2088\tablenote{side- / top-CO} & 2050 / 2080  \\
MgO(FC)-Pd-(CO$_3$)-CO &  & 
1687 / 2020\tablenote{maximum of the CO$_3$ frequencies / side-CO} &   \\
MgO(FC)-Pd$_2$-CO  & 1.95  & 1877\tablenote{bridge-bonded CO}   & 1870 \\
MgO-Pd$_2$-CO   & 2.87  & 1836$^{\rm d}$   & 1870 \\
\end{tabular}
\end{table}

%
\begin{figure}
\caption[]{
A TPR spectrum showing 
formation of CO$_2$ on a MgO(FC)-Pd sample after preadsorption of O$_2$  and
saturation with CO at 90 K.
The insets show the  infrared spectra 
of adsorbed $^{13}$CO  
after heating the sample
to the indicated temperatures. All the spectra were
recorded at 90 K.
}
\end{figure}
\begin{figure}
\caption[]{(color)
(a) Optimized structure of the MgO(FC)-Pd-(CO)$_2$O$_2$ complex. 
(b,c) selected configurations, 
and (d)  the potential energy vs time, recorded  
in  an ab-initio MD simulation 
where CO$_2$ is formed from the complex shown in (a). The simulation 
starts from the transition state shown in (b). The potential energy of
the transition state is 0.84 eV above the optimized 
configuration shown in  (a). (c) A snapshot
at 210 fs, where the formed CO$_2$ is desorbing and the remaining
O atom from O$_2$ molecule is moving towards the F-center.
The pertinent structural parameters in (a) are:
top-CO: d(C-O)=1.149 \AA, d(Pd-C)=2.10 \AA, $\angle$(Pd-C-O)=
178$^{\rm o}$; side-CO: d(C-O)=1.164 \AA, d(Pd-C)=1.91 \AA,
$\angle$(Pd-C-O)=166$^{\rm o}$; O$_2$: d(O-O)=1.46 \AA,
d(O-Pd)=2.09 \AA\  and 2.31 \AA.
The Pd atom is black, Mg and O ions of the substrate are blue
and red, respectively, adsorbed O$_2$ is yellow, carbons are 
grey and their respective oxygens are purple and 
green.
}
\end{figure}
\begin{figure}
\caption[]{(color)
(a) Optimized structure of the 
 MgO(FC)-Pd-(CO$_3$)CO complex. (b,c) selected configurations,
 and 
(d)  the potential energy vs time recorded 
in an ab-initio MD simulation 
where the carbonate complex is decomposing at around 500 K. 
Snapshots from this simulation are shown in (b) at 160 fs
(just after the transition state) and
(c) at 240 fs. The FC is filled by an
O atom in (c).
The pertinent structural parameters in (a) are:
the side CO:
d(C-O)=1.160 \AA, d(Pd-C)=1.90 \AA, $\angle$(Pd-C-O)=166$^{\rm o}$;
CO$_3$: d(Pd-O)=2.07 \AA,
 d(C-O)=1.45 \AA, 1.23\AA, 1.29 \AA\,
with the long C-O bond forming to the Pd-bound oxygen.
Colors as in Fig. 2, with the carbonate oxygens
shown in yellow.
}
\end{figure}
\end{multicols}

%
%
\end{document}